\documentclass[12pt, a4paper]{article}

\usepackage[utf8]{inputenc}
\usepackage{amsfonts,amsmath,amssymb}
\usepackage{bbold}
\usepackage{dsfont}
\usepackage{multicol}
\usepackage{cite}
\usepackage{slashed}
\usepackage{tikz}
\usepackage[compat=1.1.0]{tikz-feynman}
\usepackage{bigints}

\usepackage[
   top=10mm,
   bottom=20mm,
   left=20mm,
   right=20mm]{geometry}

\usepackage{xcolor}
\usepackage{authblk,ulem}
\usepackage{hyperref} 
\usepackage{caption}
\usepackage{graphicx}

\newcommand{\beq}{\begin{equation}}
\newcommand{\eeq}{\end{equation}}
\newcommand{\ba}{\begin{array}}
\newcommand{\ea}{\end{array}}
\newcommand{\bea}{\begin{eqnarray}}
\newcommand{\eea}{\end{eqnarray} }
\newcommand{\bal}{\begin{align}}
\newcommand{\eal}{\end{align}}

\title{Addressing $\gamma_5$ in Nondimensional Regularizations: A Case Study on the Bumblebee Model }

\author[1]{Ricardo J. C. Rosado}

\author[2]{Adriano Cherchiglia}

\author[3]{Marcos Sampaio}

\author[1]{Brigitte Hiller}

\affil[1]{CFisUC, Department of Physics, University of Coimbra, P-3004-516 Coimbra,
Portugal}
\affil[2]{Instituto de Física Gleb Wataghin, Universidade Estadual de Campinas \\ Rua Sérgio Buarque de Holanda, 777, Campinas, SP, Brasil}
\affil[3]{ Universidade Federal do ABC, 09210-580 , Santo Andr\'e, Brasil}

\begin{document}

\maketitle

\begin{abstract}
  We examine the subtleties of regularization schemes in four-dimensional space ($4S$), related in particular to the introduction of the $\gamma_5$ matrix. To illustrate we use a ``Bumblebee'' model featuring dynamically induced Lorentz symmetry violation. The analysis centers on how different regularization methods affect the solutions to the gap equation in this model. We highlight the resolution of ambiguities associated with the $\gamma_5$ matrix in ultraviolet divergent integrals by employing an enhanced Implicit Regularization (IREG) method. This method extends IREG to a quasi-four-dimensional space, $Q4S = 4S \oplus X$, drawing parallels with the consistent approach of Dimensional Reduction (DRED). Comparative analysis is conducted against results from the 't Hooft-Veltman regularization scheme, conventional IREG in strict $4S$, and sharp momentum cutoff techniques. Our results illustrate a scheme to compute $\gamma_5$ interactions in physical dimension of divergent amplitudes, confirming the approach in \cite{Bruque:2018bmy}.

\end{abstract}

\section{Introduction}

The Standard Model of Particle Physics (SM) is a successful quantum theoretical framework that describes three among four of fundamental particle interactions, namely electromagnetism, the weak 
force, and the strong nuclear force — excluding only gravity. Lorentz symmetry is at the core of the SM framework. A profound connection between Lorentz symmetry and charge, parity, and time reversal (CPT) symmetry invariance is unveiled by the CPT theorem: any Lorentz-invariant local quantum field theory with Hermitian Hamiltonian must have CPT symmetry. This is important because CPT symmetry  imposes stringent constraints on the permissible interactions and processes in nature. However, whilst an interacting theory that violates CPT
necessarily violates Lorentz invariance, it is possible to have Lorentz violation without CPT violation \cite{greenberg2002c}. On the other hand, observable manifestations of an underlying unified quantum gravity theory may present signals of  Planck scale physics associated with Lorentz symmetry breaking \cite{kostelecky2004gravity}. Indeed, various quantum-gravity approaches result in a scenario where Lorentz symmetry is broken. 

Despite the absence of a complete quantum theory of gravity, investigating Lorentz symmetry violation (LV) remains pertinent. LV carries observable implications, e.g. in cosmology and the early universe. Some theoretical frameworks propose scenarios where Lorentz symmetry is dynamically broken, potentially influencing cosmic microwave background radiation and the large-scale structure of the universe \cite{zuntz2008constraining, caloni2023probing, audren2015cosmological}. In this sense, SM Extensions (SME) serve as effective field theories models for  investigating specific instances of LV providing valuable insights for probing the fundamental nature of spacetime. In practice, this approach introduces correction terms in the Lagrangian that explicitly break Lorentz symmetry, thereby enabling the generation of measurable deviations from the Standard Model predictions. Usually such LV terms are introduced in the model through spontaneous (SSB) symmetry breaking~ 
\cite{kostelecky1999nonrelativistic, yoder2012higher, lehnert2003threshold, kostelecky2001cosmological, kostelecky2002signals, kostelecky2006sensitive, klinkhamer2011consistency, schreck2012analysis, colladay2009one, mouchrek2019erratum, maluf2014einstein, altschul2005spontaneous, kostelecky2011matter, khodadi2022probing}. 
Spontaneous Symmetry Breaking (SSB) plays a crucial role, for instance, in the unified framework of weak and electromagnetic forces. Dynamical Symmetry Breaking (DSB), which is a special case of SSB, was suggested by Bjorken in 1963 as a method for the "dynamical generation of quantum electrodynamics" (QED). This approach sought to reproduce the observable effects of standard QED without presupposing the local $U(1)$ gauge invariance. DSB is an important alternative to the Higgs mechanism and plays a central role in QCD chiral symmetry breaking.

In this contribution we study a dynamically generated Lorentz symmetry breaking model inspired by a four-fermion field model effective potential generated as a quantum correction. The minimum of the potential is determined by a tadpole finite contribution which is superficially divergent and contains $\gamma_5$ Dirac matrices. It is well known that the evaluation of divergent integrals involving $\gamma_5$ matrices potentially leads to regularization dependence. Such ambiguities in defining the $\gamma_5$ algebra across different regularizations can lead to discrepancies in the evaluation of these integrals, impacting the physical predictions.
On the other hand, regularizations that operate in the physical dimension  should, in principle, be exempt from such problems. This expectation is however too naive, as discussed in detail in~\cite{Bruque:2018bmy}, where further examples can be found in~\cite{vieira:2015fra,porto2018bose,Viglioni:2016nqc,batista:2018zxf,Cherchiglia:2021uce}. In a nutshell, it was shown in those references that identities such as $\{\gamma_{\mu},\gamma_5\}=0$ do not hold under regularization, in general. For the model studied on this work, the inability to employ this identity will require to redefine the fermionic propagator, leading to a finite, non-null result to the minimum of the model's potential. 

This work is organized as follows: after introducing the model to be used as our test case, we explore in Section \ref{sec:gamma5} the ambiguities associated with using the $\gamma_5$ matrix algebra in four-dimensional space ($4S$) with divergent integrals in IREG, and how IREG ensures result uniqueness. Section \ref{sec:reg} details the solutions to the gap equation for the Bumblebee model using IREG in both strict $4S$ and quasi-$4S$ ($Q4S$), comparing these to results from the 't Hooft-Veltman dimensional scheme~\cite{assuncao2017dynamical}, and two sharp cutoff schemes in $4S$. We present our conclusions in Section \ref{sec:con}, which are followed by three appendices: one reviews the basic rules of IREG, the second presents results for the finite integrals specific to the IREG method used in this analysis and the last discusses the global chiral symmetry of the model within IReg in Q4S.

\section{The model}
\label{sec:model}

In the late eighties, Kosteleck\'y and Samuel \cite{kostelecky1989gravitational} studied a model based on the Einstein-Maxwell action with a potential for the vector field that induces LV  when a vector (or tensor) field acquires a non-zero vacuum expectation value. For instance a potential  term  $1/(2 \alpha) (B^\mu B_\mu - \beta^2)^2$ for a background field $B_\mu$ such that $\langle 0|B_\mu|0\rangle = \beta_\mu \ne 0$  can act as indicator of global Lorentz violation. The potential reaches its minimum when the condition $ B^\mu B_\mu = \beta^2$ is met, which can be achieved, for instance, by considering a time-like 4-vector $ B_\mu = (\beta, 0, 0, 0) $. Obviously, this choice of $B_\mu$  establishes a preferred direction in spacetime.

For definiteness, consider the Bumblebee-type model given by the Lagrangian in which we have a Lagrange-multiplier potential proportional to a positive parameter $\lambda$: 
\begin{equation}
\mathcal{L}_B = -\frac{1}{4} F_{\mu\nu} F^{\mu\nu} + \bar{\psi}(i\partial_\mu \gamma^\mu - m - e \slashed{B}\gamma_5)\psi - \frac{\lambda}{4}(B_\mu B^\mu - \beta^2)^2.
\label{eq:BB}
\end{equation}
Here $F_{\mu\nu} = \partial_\mu B_\nu - \partial_\nu B_\mu$. Performing a shift $B_\mu \rightarrow \beta_\mu + A_\mu$, with $\langle A_\mu \rangle = 0$, yields: 
\begin{equation}\label{eq:bumbe}
\mathcal{L}_B = -\frac{1}{4} F_{\mu\nu} F^{\mu\nu} + \bar{\psi}[i\slashed{\partial} - m -  (e\slashed{A} + \slashed{b})\gamma_5]\psi - \frac{\lambda}{4} \left( A_\mu A^\mu + \frac{2}{e} A_\mu b^\mu \right)^2,
\end{equation}
in which the term in $b_\mu=e\beta_\mu$ explicitly violates Lorentz symmetry. We remark that the present model does not possess gauge symmetry either. However, it does comply with global chiral symmetry. This  symmetry is trivially fulfilled in the bosonization procedure $B_\mu \sim \bar\psi \gamma_\mu \gamma_5 \psi$ considered below, as $B_\mu$ inherits the chiral symmetry invariance of the fermion bilinear of the underlying four fermion interaction Lagrangian, eq. \ref{fourferm}. By the same token chiral symmetry persists through the shift $B_\mu=A_\mu +\beta_\mu$. An interesting question is if this chiral symmetry still persists after regularization. We will discuss this issue in Appendix \ref{sec:symmetry} for the case of IREG.

A different venue to arrive at this model comes by following the work of Coleman and Weinberg \cite{coleman1973radiative} that established that spontaneous symmetry breaking can be generated at quantum level. It has been demonstrated in the literature that Bumblebee-type potentials, analogous to those found in the Bumblebee model of gauge symmetry breaking, can arise dynamically within certain theoretical frameworks. \cite{gomes2008dynamical}. In particular, considering as a starting point an interaction term consisting of a fermion  bilinear of massless fields transforming as an axial-vector under Lorentz transformations \cite{gomes2008dynamical}:
\begin{eqnarray}
\label{fourferm}
\mathcal{L} &=& \bar{\psi} i \slashed \partial \psi - \frac{e^2}{2g^2} (\bar{\psi} \gamma_\mu \gamma^5 \psi)^2 +
\frac{g^2}{2} \left[ B_\mu - \frac{e}{g^2} \bar{\psi} \gamma_\mu \gamma^5 \psi \right]^2\nonumber \\
&=&  \frac{g^2}{2} B_\mu B^\mu + \bar{\psi} (i \slashed \partial - e \slashed B \gamma^5) \psi,
\end{eqnarray}
in which the term $- \frac{e^2}{2g^2}(\bar{\psi} \gamma_\mu \gamma^5 \psi)^2 $ has been cancelled out. Following \cite{assuncao2017dynamical}, after some manipulations we arrive at the effective potential for the Bumblebee field:
\begin{equation}
    V_{\text{eff}} = - \frac{g^2}{2} B_\mu B^\mu + i \text{tr} \int \frac{d^4k}{(2\pi)^4} \ln(\slashed{k} - e \slashed B \gamma^5).
\end{equation}
The nontrivial minimum of this potential is obtained by imposing:
\begin{equation}
\frac{dV_{\text{eff}}}{dB_\mu} \Bigg|_{e B_\mu=b_\mu} = -\frac{g^2}{e} b^\mu - i e \Pi^\mu = 0,
\end{equation}
where 
\begin{equation}
\Pi^\mu = \text{tr} \int \frac{d^4k}{(2\pi)^4} \frac{1}{\slashed{k} - \slashed{b}\gamma^5} \gamma^\mu \gamma^5
\label{eq:Pi}
\end{equation}
is the one-loop tadpole amplitude. It is evaluated in Ref. \cite{assuncao2017dynamical}, employing dimensional regularization in the Breitenlohner, Maison, ´t Hooft and Veltman (BMHV) scheme, to be discussed in the next section, to
\begin{equation}
\Pi^\mu = \frac{i b^2}{3\pi^2} b^\mu,
\end{equation}
leading to a gap equation:
\begin{equation}
\frac{dV_{\text{eff}}}{dB_\mu} \Bigg|_{eB_\mu=b_\mu} = \left(- \frac{1}{G} + \frac{b^2}{3\pi^2} \right) eb^\mu = 0.
\end{equation}
It has the non-trivial solution
\begin{equation}
b^2 = \frac{3 \pi^2}{G},
\end{equation}
with $G=\frac{e^2}{g^2}$ and $G>0$ ($G<0$) for timelike (spacelike) $b_\mu$. On the other hand, the effective potential reads
\begin{equation}\label{eq:general}
V_{\text{eff}} = -\frac{e^2b^2}{6\pi^2} B^2 + \frac{e^4}{12\pi^2} B^4 + c,
\end{equation}
where $c$ is an integration constant. In this way, the Bumblebee potential in equation \ref{eq:BB} is reproduced by choosing $c=b^4/(12 \pi^2)$ and $\lambda=e^4/(3 \pi^2)$.

This illustrates that within the four-fermion model, a bumblebee potential featuring nontrivial minima can arise due to quantum corrections, which in turn leads to dynamical Lorentz symmetry breaking within this context.
To ensure the framework's consistency, it is necessary for $\Pi^\mu$ to be finite, as this condition is essential for obtaining a physical solution that corresponds to the potential's minimum.

In this work, we explore how the coefficients in eq.~\eqref{eq:general} depend on a 
%
coherent treatment of the $\gamma_5$ algebra by 
various regularization techniques.
Our primary attention is on regularizations that are applied in the  physical dimension, as these can notably simplify higher-order calculations if a consistent approach to chiral theories is adopted.  

\section{$\gamma_{5}$ matrix algebra in implicit regularization}
\label{sec:gamma5}

Handling the $\gamma_5$ Dirac matrix in conventional dimensional regularization (CDR) is problematic, as expanding its specific algebra to arbitrary dimensions $D$ breaks chiral symmetry. To resolve this, authors have devised new schemes for precise calculations beyond leading order \cite{veltman1972regularization, breitenlohner1977dimensionally, jones1982dimensional, korner1992practicable, kreimer1994role, larin1993renormalization, schubert1993gamma_, trueman1995spurious, jegerlehner2001facts, tsai2009advantage, tsai2011gauge, ferrari2014managing, ferrari2017managing, moch2015gamma5, ferrari2016gamma_5,Chen:2023lus}, redefining algebraic rules for extended Lorentz tensors and gamma matrices, enforcing constraints order by order, and adding counterterms to restore symmetries according to quantum action principles \cite{stockinger2005regularization}.
  
On the other hand, refining theoretical predictions in particle physics requires effective computational methods for calculating Feynman amplitudes beyond the next-to-leading order (NLO). For some processes, high-precision measurements already call for theoretical calculations up to at least four loop order \cite{das2020approximate} and serve to probe extensions of the Standard Model. In this sense exploring non-dimensional regularization schemes offers a promising approach to simplify computations following some already successful mixed schemes such as dimensional reduction (DRED)~\cite{siegel1979supersymmetric,siegel1980inconsistency} and four-dimensional helicity (FDH)~\cite{bern1992computation,bern2002supersymmetric}. Among such schemes, notable examples include four-dimensional formulation (FDF) \cite{fazio2014four}, four dimensional regularization (FDR) \cite{pittau2012four}, the four dimensional unsubtracted (FDU)
method~\cite{Hernandez-Pinto:2015ysa,Sborlini:2016hat} and implicit regularization (IREG) \cite{battistel1998consistency,scarpelli2001chiral,scarpelli2001consistency,cherchiglia2011systematic}, see~\cite{gnendiger2017d,torres2021may} for reviews.

In conventional dimensional methods (CDR), a primary challenge comes from the relation between the emergence of anomalous terms and the breakdown of cyclicity — a characteristic typically preserved in finite dimensions \cite{kreimer1990gamma5}. While in CDR $\{\gamma_{5},\gamma_\mu\} = 0$ is typically preserved in $D\ne 4$, in the BMHV extension  \cite{breitenlohner1977dimensionally} $\{\gamma_{5},\gamma_\mu\} \ne 0$  and, as a byproduct, gauge and BRST invariance should be consistently restored by symmetry-restoring counterterms at all orders \cite{belusca2020dimensional,belusca2021two,Stockinger:2023ndm}. One might anticipate that for four-dimensional regularization schemes, the $\gamma_5$ matrix issue would not arise. Nonetheless, there exist subtleties related to finite ambiguities when integrating over internal momenta. One example is symmetric integration in the internal loop momenta \cite{perez2001physical}. Such operation alters the Lorentz structure and consequently the $\gamma_5$ Clifford algebra by introducing spurious terms \cite{porto2018bose,Bruque:2018bmy,Viglioni:2016nqc,Cherchiglia:2021uce}. In a nutshell, the contraction of Lorentz indices does not commute with renormalization in these non-dimensional schemes. To circumvent this problem, Bruque and collaborators \cite{Bruque:2018bmy}  proposed a scheme in which Dirac matrices, with the exception of $\gamma_5$, are defined in a quasi-dimensional space $Q4S = QdS\oplus Q(2\epsilon)S$ \cite{stockinger2005regularization}. 
However, in contrast to DRED (where the momenta still need to be treated in $QdS$) the amplitude momenta are also defined in $Q4S$. One defines $Q4S= 4S \oplus X$, $X$ being an auxiliary space, which needs not be explicitly defined \cite{Bruque:2018bmy}. In the specific case of IREG, the inconsistencies boil down to the contraction of internal momenta in Feynman amplitudes. Let us illustrate this point with a toy integral\footnote{A concise summary of IREG's fundamental rules can be found in the appendix.} involving the internal momentum $k$ and symmetric integration: $k^2 \rightarrow k_\alpha k_\beta g^{\alpha \beta}$ or $k_\alpha k_\beta \rightarrow \frac{k^2}{4} g_{\alpha \beta}$. On one hand,

\begin{equation}
\int_{k}  \dfrac{k^{2}}{k^{2}(k-p)^{2}}=\int_{k}\dfrac{1}{(k-p)^{2}}=\lim_{\mu^{2}\rightarrow 0}\int_{k}  \dfrac{1}{(k-p)^{2}-\mu^{2}}=\lim_{\mu^{2}\rightarrow 0} \int_{k}  \dfrac{1}{k^{2}-\mu^{2}}=0
\label{Nk2}
\end{equation}
whereas, performing symmetric integration,
\begin{small}
\begin{equation}
\begin{aligned}
g^{\alpha\beta}\int_{k}  \dfrac{k_{\alpha}k_{\beta}}{k^{2}(k-p)^{2}}
&=g^{\alpha\beta}\left\{\left(\frac{p_{\alpha}p_{\beta}}{3}-\frac{g_{\alpha\beta}p^{2}}{12}\right)\left[I_{\text{log}}(\lambda^{2})- b \ln\left(-\frac{p^2}{\lambda^2}\right) + \frac{13b}{6}\right]-\frac{g_{\alpha\beta}b p^{2}}{24}\right\}\\
&=-\frac{b}{6}p^{2},
\label{gab_kakb}
\end{aligned}
\end{equation}
\end{small}

\noindent
where $b= i/(4\pi)^2$, and $\int_k \equiv \int d^4k/(2 \pi)^4$, clearly showing an ambiguity in the evaluation, which is dependent on regularization since symmetric integration is not applicable in general \cite{perez2001physical}. In the $\gamma_5$ matrix algebra, such ambiguity  emerges in non-dimensional regularization schemes  upon consistently requiring  $\{\gamma_{\mu},\gamma_{5}\}=0$:
\begin{equation}
\int_{k}  \dfrac{\slashed{k}\gamma_{5}\slashed{k}}{k^{2}(k-p)^{2}}=
\begin{cases}
0,& \text{using eq. (\ref{Nk2})\, or}\\
\frac{b}{6}p^2\gamma_{5}, & \text{using eq.(\ref{gab_kakb}) }.
\end{cases}
\label{eqn:gamma5}
\end{equation}

In order to avoid these ambiguities, one defines
\begin{equation}
\gamma_{5}=-\frac{i}{4!}\epsilon_{abcd}\bar{\gamma}^{a}\bar{\gamma}^{b}\bar{\gamma}^{c}\bar{\gamma}^{d}
\label{eq:def G5}    
\end{equation}
where we use an overbar to denote an object pertaining to $4S$. Since the Dirac matrices are defined in $Q4S$, they are required to satisfy:
\begin{align}\label{eq:comute}
\{\gamma_{\mu},\gamma_{\nu}\}&=2g_{\mu\nu}\mathbb{1}; \quad \{\bar{\gamma}_{\mu},\bar{\gamma}_{\nu}\}=\{\gamma_{\mu},\bar{\gamma}_{\nu}\}=2\bar{g}_{\mu\nu}\mathbb{1}; \quad \gamma_{\mu}\gamma^{\mu}=\gamma_{\mu}\bar{\gamma}^{\mu}=4\;\mathbb{1}\\
\{\bar{\gamma}_{\mu},\hat{\gamma}_{\nu}\}&=0; \quad \{\gamma_{\mu},\hat{\gamma}_{\nu}\}=\{\hat{\gamma}_{\mu},\hat{\gamma}_{\nu}\}=2\hat{g}_{\mu\nu}\mathbb{1};\quad \gamma_{\mu}\hat{\gamma}^{\mu}=\bar{\gamma}_{\mu}\hat{\gamma}^{\mu}=\hat{\gamma}_{\mu}\hat{\gamma}^{\mu}=0;\\
\{\bar{\gamma}_{\mu},\gamma_{5}\}&=0; \quad \{\gamma_{\mu},\gamma_{5}\}=2\gamma_{5}\hat{\gamma_{\mu}}; \quad [\hat{\gamma}_{\mu},\gamma_{5}]=0
\label{eq:comute2}
\end{align}
where hatted objects belong to $X$ space. In view of the above properties, equation (\ref{eqn:gamma5}) is evaluated as \cite{Cherchiglia:2021uce}
\begin{align}
\int_{k}  \dfrac{\slashed{k}\gamma_{5}\slashed{k}}{k^{2}(k-p)^{2}}&=
2\gamma_{5}\int_{k}  \dfrac{\hat{\slashed{k}}\slashed{k}}{k^{2}(k-p)^{2}}-\int_{k}  \dfrac{\gamma_{5}k^2}{k^{2}(k-p)^{2}}\nonumber\\
&=\gamma_{5}\int_{k}  \dfrac{k^2}{k^{2}(k-p)^{2}}-2\gamma_{5}\int_{k}  \dfrac{\bar{k}^2}{k^{2}(k-p)^{2}}\nonumber\\
&=-2\bar{g}_{ab}\gamma_{5}\int_{k}  \dfrac{\bar{k}^{a}\bar{k}^{b}}{k^{2}(k-p)^{2}}=\frac{bp^{2}}{3}\gamma_{5}. \label{eq:third}
\end{align}

Notice that the above result differs from the two naive options of equation~\ref{eqn:gamma5}. Since this is just a toy integral, it is not possible to connect any of the approaches to the (possible) breaking of Ward identities. However, only the last approach, extending the theory to the $Q4S$ space, is completely free of ambiguities (particularly at multiloop calculations), since it does not require reading points to be defined \cite{tsai2009advantage, tsai2011gauge,tsai2011maintaining}, for instance \footnote{The main drawback is that chiral symmetry is inevitably broken, as happens in the BMHV scheme. It can, nevertheless, be restored by including finite counterterms (see \cite{Bruque:2018bmy,Cherchiglia:2021uce} for examples in IREG).}.

\section{The bumblebee model from distinct regularizations}
\label{sec:reg}

In \cite{assuncao2017dynamical}, $\Pi^\mu$ in equation \ref{eq:Pi} was evaluated by employing the exact propagator and using  dimensional regularization with ’t Hooft-Veltmann prescription \cite{veltman1972regularization} following an analytical extension from 4 to a $D$-dimensional spacetime. Moreover Dirac matrices are required to obey
 $\{{\gamma}^\mu, {\gamma}^\nu\} = 2{g}^{\mu\nu}$, with  ${g}_{\mu\nu}{g}^{\mu\nu} = D$. 
 Dirac matrices ${\gamma}^\mu$ and the metric tensor ${g}^{\mu\nu}$ are split as 
\begin{eqnarray} 
{\gamma}^\mu &=& \bar{\gamma}^\mu + \hat{\gamma}^\mu, \nonumber\\ 
{g}^{\mu\nu} &=& \bar{g}^{\mu\nu} + \hat{g}^{\mu\nu}, 
\end{eqnarray}
that is, into 4-dimensional parts (expressed with a bar) and $(D - 4)$-dimensional parts (hatted), so that now the Dirac matrices satisfy the relations
\begin{equation}
\{ \bar{\gamma}^\mu, \bar{\gamma}^\nu \} = 2\bar{g}^{\mu\nu}, \quad \{ \hat{\gamma}^\mu, \hat{\gamma}^\nu \} = 2\hat{g}^{\mu\nu}, \quad \{ \bar{\gamma}^\mu, \hat{\gamma}^\nu \} = 0,
\end{equation}
and the metric tensors obey 
\begin{equation}
\bar{g}_{\mu\nu}\bar{g}^{\mu\nu} = 4, \quad \hat{g}_{\mu\nu}\hat{g}^{\mu\nu} = D - 4,
\quad  \bar{g}_{\mu\nu}\hat{g}^{\mu\nu} = 0. 
\end{equation}
In this way one has
\begin{equation}
[\hat{\gamma}^\mu, \gamma^5] = 0
\end{equation}
and  the anticommutation relation
\begin{equation}
\{ \bar{\gamma}^\mu, \gamma^5 \} = 0.
\end{equation}
is maintained. 

We proceed to study how eq.\eqref{eq:Pi} can be evaluated in four-dimensional regularization schemes, such as IREG. Before discussing the gap equation, though, we will take a step back and discuss how the Feynmann rules are extracted from eq.\eqref{eq:bumbe}. 
This task may be non trivial in the presence of $\gamma_5$, which is only defined in the physical dimension. In particular, if the regularization to be adopted is defined in a different dimension, the Lagrangian (and thereof the Feynman rules) must also be extended to this dimension. Once $\gamma_{\mu}$ is defined in the extended dimension, the property $\{\gamma_{\mu},\gamma_{5}\}=0$ may not hold. For the bumblebee model this implies that the fermionic propagator ($D_{B}$) will not be written in terms of the left/right propagators ($P_{L}/P_{R}$) only. Consider, for instance, that 
$\{\gamma^\mu,\gamma^5\}=2 \hat{\gamma}^\mu \gamma^5$, where $\hat{\gamma}^\mu$ is defined in the extra space $X$ defined in the previous section. At first sight, we have 
\begin{equation}
    D_B= \frac{1}{\slashed{k}-\slashed{b}\gamma^5},\quad D_B^{-1}=\left( (\slashed{k}-\slashed{b}) P_R + (\slashed{k}+\slashed{b}) P_L \right)\;.
\end{equation}
However, it is straightforward to show that
\begin{equation}
    D_B  D^{-1}_B = 1 - \left( \frac{1}{\slashed{k}-\slashed{b}} + \frac{1}{\slashed{k}+\slashed{b}} \right) (\slashed{\hat{k}}-\slashed{\hat{b}}\gamma^5)\;.
\end{equation}
Thus, $D_B^{-1}$ cannot be the inverse propagator, unless $\{\gamma^\mu,\gamma^5\}=0$ holds.

In the following, consider that the regularization to be employed is defined in the physical dimension, implying that $\{\gamma^\mu,\gamma^5\}=0$, which allows to write $\Pi^\mu$ as
\begin{align}\label{eq:gap}
    \Pi^\mu = Tr\left[ \int_k \frac{1}{\slashed{k}-\slashed{b}\gamma^5} \gamma^\mu \gamma^5 \right] &=  Tr\left[ \int_k \left( \frac{1}{\slashed{k}-\slashed{b}} P_L + \frac{1}{\slashed{k}+\slashed{b}} P_R \right) \gamma^\mu \gamma^5 \right] \nonumber\\
    &= 2 \int_k \frac{k^\mu-b^\mu}{(k-b)^2} - 2 \int_k \frac{k^\mu+b^\mu}{(k+b)^2}\nonumber\\
    &=2 \int \frac{k^\mu[(k+b)^2-(k-b)^2]-b^\mu[(k+b)^2+(k-b)^2]}{(k-b)^2(k+b)^2}    
\end{align}

After employing Feynman parametrization, we obtain
\begin{align}\label{eq:phys}
    \Pi^\mu = Tr\left[ \int_k \frac{1}{\slashed{k}-\slashed{b}\gamma^5} \gamma^\mu \gamma^5 \right]  =-2 \int^1_0 dx \int_{k} \frac{2b^\mu(k^2-\Delta)-4k^\mu(b \cdot k)}{(k^2-\Delta)^2} 
\end{align}
where $\Delta=4b^2x(x-1)+\mu^2$, and $\mu^{2}$ is a fictitious mass introduced in the propagators of eq.\ref{eq:gap}. This result should be compared with the one obtained in the framework of Dimensional Regularization in the BMHV scheme~\cite{assuncao2017dynamical}
\begin{equation}
\label{trapi}
     Tr\left[ \int_k \frac{1}{\slashed{k}-\slashed{b}\gamma^5} \gamma^\mu \gamma^5 \right]  = -2 \mu^{4-D} \int^1_0 dx \int \frac{dk^{d}}{(2\pi)^{d}}\frac{2b^\mu(k^2-M^2-2\hat{k}^2)-4k^\mu(b \cdot k)}{(k^2-M^2)^2} 
\end{equation}
where $M^{2}=4b^2x(x-1)$. The main difference is the appearance of a term containing $\hat{k}^2$.

\subsection{Naive IREG approach}

We proceed to compute eq.\ref{eq:phys} using IREG. The main idea is to extract the UV divergent part, in a way that the integrals only depend on internal momenta. This can be achieved employing the following identity as many times as necessary
\begin{equation}
    \frac{1}{k^2-M^2-\mu^2} = \frac{1}{k^2-\mu^2}\left(1+\frac{M^2}{k^2-M^2-\mu^2}.\right)
\end{equation}

For instance, 
\begin{align}
\label{eq:I2}
      I_{2}&=\int_k \frac{1}{k^2-\Delta} = \underbrace{ \int_k  \frac{1}{k^2-\mu^2}}_{\text{Quad Div}} + \underbrace{ \int_k  \frac{M^2}{(k^2-\mu^2)^2}}_{\text{Log Div}} + \underbrace{ \int_k  \frac{M^4}{(k^2-\mu^2)^2(k^2-M^2-\mu^2)}}_{\text{Finite}}\\
\label{eq:I2m}      
      I_{2}^{\mu\nu}&=\int_k\frac{k^\mu k^\nu}{(k^2-\Delta)^2} =\underbrace{\int_k \frac{k^\mu k^\nu}{(k^2-\mu^2)^2}}_{\text{Quad Div}} + \underbrace{\int_k \frac{2 k^\mu k^\nu M^2}{(k^2-\mu^2)^3}}_{\text{Log Div}} +\nonumber\\
&\quad\quad\quad\quad\quad\quad\quad\quad+ \underbrace{\int_k \frac{2 k^\mu k^\nu M^4}{(k^2-\mu^2)^3(k^2-M^2-\mu^2)}}_{\text{Finite}} + \underbrace{\int_k\frac{k^\mu k^\nu M^4}{(k^2-\mu^2)^2(k^2-M^2-\mu^2)^2}}_{\text{Finite}}
\end{align}

In terms of the integrals above one obtains
\begin{align}
\label{pint}
    \Pi^\mu = Tr\left[ \int_k \frac{1}{\slashed{k}-\slashed{b}\gamma^5} \gamma^\mu \gamma^5 \right]  =-2 \int^1_0 dx \left(2b^\mu I_{2}-4b_{\alpha} I_{2}^{\mu\alpha}\right)\,.
\end{align}
By employing the identities
\begin{align}\label{eq:TS}
  \int_k \frac{k_\mu k_\nu}{(k^2-\mu^2)^2} = \frac{g_{\mu\nu}}{2}\int_k \frac{1}{(k^2-\mu^2)}\,,\quad\quad \int_k \frac{k_\mu k_\nu}{(k^2-\mu^2)^3} = \frac{g_{\mu\nu}}{4}\int_k \frac{1}{(k^2-\mu^2)^2}\,,
\end{align}
it is immediate to notice that all divergences will cancel in the gap equation. Regarding the finite pieces, they will also cancel, as shown in appendix \ref{finite}.
Therefore, if the regularization is defined in the physical dimension (in particular if $\{\gamma^\mu,\gamma^5\}=0$ does hold), we obtain a null result for $\Pi^\mu$, in disagreement with \cite{assuncao2017dynamical}.

There are, however, subtleties when defining a regularization in the physical dimension. In particular, if the regularization complies with shift invariance, it can be shown \cite{Bruque:2018bmy} that $f(\{\gamma^\mu,\gamma^5\})I_{\mu\nu}\neq0$. Here $f()$ stands for an expression containing Dirac matrices (for instance a Dirac Trace), and $I_{\mu\nu}$ is a divergent integral. The crucial point is that both indexes of the integral $I_{\mu\nu}$ are contracted when multiplying $f()$. From a more formal point of view, the fact that $\{\gamma^\mu,\gamma^5\}=0$ does not hold under regularization can be incorporated by extending the dimension of the underlying Lagrangian, while $\gamma_5$ is still only defined in the physical dimension. In  the framework of Dimensional Regularization, this stands for the BHMV approach \cite{belusca2020dimensional, belusca2021two,Stockinger:2023ndm} For IREG, a similar approach can be envisaged \cite{Bruque:2018bmy}. We will employ this approach in the next section.

\subsection{IREG in Q4S}

Once we define the theory on Q4S, we can treat the propagator of the Bumblebee theory (eq.~\ref{eq:bumbe}), which is given by
\begin{align}
    D_B= \frac{1}{\slashed{k}-\slashed{b}\gamma^5}.
\end{align}
Notice that the $\slashed{k}$, $\slashed{b}$ were extended to Q4S, while $\gamma_{5}$ stays in 4S. Using the properties regarding Dirac matrices in Q4S (see eqs.\eqref{eq:comute}-\eqref{eq:comute2}),
it can be shown that the propagator may also be expressed as
\begin{equation}
\label{prop}
     D_B= \frac{1}{\slashed{k}-\slashed{b}\gamma^5} = \frac{k^2 +\overline{b}^2 + (2 k \cdot \overline{b} + [\slashed{\hat{k}},\slashed{\bar{b}}]) \gamma^5}{(k-\overline{b})^2 (k+\overline{b})^2 - 4 \hat{k}^2 \overline{b}^2} (\slashed{k}+\slashed{\bar{b}}\gamma^5)\;.
\end{equation}

Thus, the tadpole contribution to the gap equation is now obtained as 
\begin{align}
    \Pi^\mu =  Tr\left[ \int_k \frac{1}{\slashed{k}-\slashed{b}\gamma^5} \gamma^\mu \gamma^5 \right] = Tr\left[ \int_k \frac{k^2 +\overline{b}^2 + (2 k \cdot \overline{b} + [\slashed{\hat{k}},\slashed{\bar{b}}]) \gamma^5}{(k-\overline{b})^2 (k+\overline{b})^2 - 4 \hat{k}^2 \overline{b}^2} (\slashed{k}+\slashed{\bar{b}}\gamma^5) \gamma^\mu \gamma^5 \right]\;.
\end{align}

After a tedious, yet straightforward calculation, we obtain
\begin{equation}
\label{eq:ag}
    \Pi^\mu (b) = - 2 \int^1_0 dx \int_{k} \frac{2\overline{b}^\mu(k^2-\Delta) -4 k^\mu \left(\overline{b}  \cdot k\right)}{(k^2-\Delta)^2} + 8 \int^1_0 dx \int_{k} \frac{\overline{b}^\mu \hat{k}^2}{(k^2-\Delta)^2}  
\end{equation}
which should be compared against eq.\ref{eq:phys}. The only difference in the first term of eq. \ref{eq:ag} as compared to the result written in the physical dimension (eq.\ref{eq:phys}) is that the internal momentum is now defined in Q4S. 
However, all the steps leading to the integrals $I_{2}^{\mu\nu}$ and $I_{2}$ still hold. Therefore, we arrive at the same conclusion as before, the first term is null, and one gets simply 
\begin{equation}
    \Pi^\mu (b) =  8 \int^1_0 dx \int_{k} \frac{\overline{b}^\mu \hat{k}^2}{(k^2-\Delta)^2}\;.
\end{equation}

In order to evaluate this integral, we notice that $\hat{k}^2=k^2-\bar{k}^2$. In the framework of IREG, we have the property below (required to fulfill shift invariance\footnote{We recall that, if a regularization method complies with shift invariance, Ward identities related to abelian gauge symmetry are automatically respected (see \cite{ferreira:2011cv,Viglioni:2016nqc} for further discussion of this point in the context of IREG).} in Q4S) \cite{Bruque:2018bmy}
\begin{align}
    \int_{k} k^{2}f(k) \neq g_{\alpha\beta} \int_{k} k^{\alpha}k^{\beta} f(k)\;. 
\end{align}
This ultimately implies that Lorentz contraction and regularization do not commute, once the internal momenta are in Q4S. On the other hand, for contracted internal momenta in 4S, we have
\begin{align}
\label{cont4S}
    \int_{k} \bar{k}^{2}f(k) = \bar{g}_{\alpha\beta} \int_{k} k^{\alpha}k^{\beta} f(k)\;. 
\end{align}

Thus, the gap equation can be expressed in terms of 
\begin{equation}
    \Pi^\mu (b) =  8\overline{b}^\mu \left[ \int^1_0 dx \int_{k} \frac{ k^2}{(k^2-\Delta)^2} - \bar{g}_{\alpha\beta} \int^1_0 dx \int_{k} \frac{k^{\alpha}k^{\beta}}{(k^2-\Delta)^2}\right]\;.
\end{equation}
As before these integrals should be decomposed in its finite and basic divergent contributions. With $N(k)=k^2 -\bar{g}_{\alpha\beta}k^{\alpha}k^{\beta}$ and recalling that $\Delta=M^{2}+\mu^{2}$  one has
\begin{align}
       &\int_k\frac{N(k)}{(k^2-\Delta)^2} =\underbrace{\int_k \frac{N(k)}{(k^2-\mu^2)^2}}_{\text{Quad Div}} + \underbrace{\int_k \frac{2  M^2 N(k)}{(k^2-\mu^2)^3}}_{\text{Log Div}} +  \underbrace{\int_k \frac{2  M^4 N(k)}{(k^2-\mu^2)^3(k^2-M^2-\mu^2)}}_{\text{Finite}}\nonumber\\
&\quad\quad\quad\quad\quad\quad\quad\quad + \underbrace{\int_k\frac{M^4 N(k)}{(k^2-\mu^2)^2(k^2-M^2-\mu^2)^2}}_{\text{Finite}}.
\end{align}
For the finite pieces there is no distinction between $k^{2}$ and $\bar{k}^{2}=\bar{g}_{\alpha\beta}k^{\alpha}k^{\beta}$. Therefore, $N(k)$ is identically null for these terms. 

Regarding the divergent pieces one uses $N(k)=(k^2-\mu^2) +\mu^2-\bar{g}_{\alpha\beta}k^{\alpha}k^{\beta}$ to cancel powers in the denominator and be able to use the relations \ref{eq:TS}, obtaining
\begin{eqnarray}
    \Pi^\mu (b)|_{div} &=&  8\overline{b}^\mu \int^1_0 dx \left[ \int_k \frac{\mu^{2}}{(k^2-\mu^2)^2} - \int_k \frac{1}{(k^2-\mu^2)} \right]\ \\
   &+& 8\overline{b}^\mu \int^1_0 dx\int_k \frac{2 M^2 \mu^{2}}{(k^2-\mu^2)^3}. \\
\end{eqnarray}  

After taking the limit $\mu^{2}\rightarrow 0$, the first two terms are null and the last term yields the finite contribution to the gap equation, 

\begin{equation}
\label{tHV}
    \Pi^\mu (b) =  8\overline{b}^\mu \left[\frac{2i}{(4\pi)^2}\frac{b^{2}}{3}\right]=i\frac{\overline{b}^\mu}{3\pi^2}b^2\;.
\end{equation}

\subsection{Cut-off regularizations}
A widely used regularization procedure consists in applying a sharp $4D$ momentum cutoff $\Lambda$ on the divergent integrals. Here we will discuss two types of cutoff regularization. 

First we consider the standard approach, in which the tensor structures involving the loop momentum $k$ are dealt with using symmetric integration. In the case of two Lorentz indices this  corresponds to the replacement

\begin{equation} 
\label{sym}
\int_k \frac{k_\mu k_\nu}{(k^2-M^2)^n}\rightarrow  \frac{g_{\mu\nu}}{4}\int_k \frac{ k^2}{(k^2-M^2)^n}
\end{equation}
in finite as well as divergent integrals. Here $\int_k= \int \frac{d^4k}{(2 \pi)^4}$ and $M^2$ stands for any scalar dependence on momenta (other than the loop momentum), masses or Feynman parameters. In general this reduction turns out to induce all sorts of violations of symmetries when the substitution is done in divergent integrals. 

In the case of the Bumblebee model studied here the integrals to be evaluated, after Dirac trace is taken using the $\gamma_5$ algebra in $4D$ and after Feynman parametrization, are identical to  the ones of eq. \ref{trapi}, but now evaluated with a sharp cutoff $\Lambda$ instead.  With
\begin{equation}
{\int_k}^\Lambda \frac{1}{k^2-M^2}=-\frac{i}{(4 \pi)^2} (\Lambda^2 -M^2 \mbox{ln}(\frac{\Lambda^2+M^2}{M^2})),
\end{equation}
\begin{equation}
{\int_k}^\Lambda \frac{1}{(k^2-M^2)^2}=\frac{i}{(4 \pi)^2} (\mbox{ln}(\frac{\Lambda^2+M^2}{M^2}) -\frac{\Lambda^2}{\Lambda^2+M^2}),
\end{equation}
a simple calculation using the symmetric integration (\ref{sym})
leads to the result
\begin{equation}
\label{symPi}
\Pi_\mu=\frac{i b_\mu}{4}\left(\frac{b^2}{3\pi^2}+\frac{\Lambda^2}{2 \pi^2}\right).
\end{equation}
This result differs from the `t Hooft Veltman scheme, eq. \ref{tHV}, by a cutoff dependence, which antagonizes with the effective potential requirement for a finite result. Interestingly the finite term has also a different coefficient. 

Secondly we consider the gauge invariant sharp cutoff procedure of \cite{cynolter2015cutoff}. 
In this particular calculation the main difference to the naive procedure just outlined is that the coefficient that accompanies the reduction to the metric tensor in divergent integrals  results from the requirement that the surface terms (ST) relating the difference of the following two quadratic divergences, or the following two logarithmic divergences, vanish 
\begin{equation}
\label{ST1}
{\int_k} \frac{\partial}{\partial k_\nu} \frac{k^\mu}{k^2-M^2} =g^{\mu\nu}{\int_k} \frac{1}{k^2-M^2}-2{\int_k} \frac{k^\mu k^\nu}{(k^2-M^2)^2} =0, 
\end{equation}
\begin{equation}
\label{ST2}
{\int_k} \frac{\partial}{\partial k_\nu} \frac{k^\mu}{(k^2-M^2)^2}=g^{\mu\nu}{\int_k} \frac{1}{(k^2-M^2)^2}-4{\int_k}^\Lambda \frac{k^\mu k^\nu}{(k^2-M^2)^3} =0.
\end{equation}
The vanishing of ST complies with the requirement of momentum routing invariance (the invariance under shifts in the loop momentum) and is at the core of gauge invariance (in the case of IREG the  conditions are embodied in eqs. \ref{eq:TS}).  In order to be able to relate to a simple momentum cutoff, the authors identify a set of rules \cite{cynolter2015cutoff} and get
\begin{equation}
\label{tensor}
{\int_k}^\Lambda \frac{k^\mu k^\nu}{(k^2-M^2)^2}=- \frac{g^{\mu\nu}}{2}\frac{i}{(4 \pi)^2} \left[\Lambda^2 -M^2 \mbox{ln}\left(\frac{\Lambda^2+M^2}{M^2}\right)\right].
\end{equation}
One finally obtains that the improved (gauge invariant) sharp cutoff momentum prescription leads to an identical result as in IREG in strict four-dimensions, namely a null result for $\Pi_\mu$.

To summarize, the gap equation of the Bumblebee model depends on the regularization employed. Resorting to naive gauge invariant regularizations, one obtains a null result (naive DReg, naive IReg and gauge invariant sharp cut-off). However, care must be exercised since we have a chiral vertex in the theory. Therefore, the regularization can only be consistently defined in a quasi-dimensional space (Q4S), implying that identities only applicable at the genuine physical dimension do not hold anymore. In particular, the property $\{\gamma_{\mu},\gamma_{5}\}=0$ cannot be used, leading to a non-null result for the gap equation when employing the consistent version of IReg or DReg in the BHMV scheme.

\section{Conclusions}
\label{sec:con}
 The development of regularization methods that operate entirely or partially in the physical dimension aims to automate calculations at and beyond next-to-leading order (NLO). Traditional dimensional regularization schemes often increase complexity when handling objects like the $\gamma_5$ matrix, which are well-defined only in the physical dimension. This complexity has motivated the adoption of non-dimensional methods. However, it's recognized that not all operations of the $\gamma_5$ algebra are applicable to divergent integrals without introducing ambiguities. In the case of Implicit Regularization (IREG), inconsistencies at NLO can generally be resolved by either symmetrizing the trace in divergent integrals \cite{vieira:2015fra,Viglioni:2016nqc,porto2018bose} or applying the ``right-most position'' technique in open fermionic strings \cite{batista:2018zxf,rosado2023infrared}, both maintaining strict adherence to the physical dimension.
In this study, a new layer of complexity is introduced as the $\gamma_5$ matrix also appears in the fermionic propagator. Its handling in the physical dimension involves using $\{\gamma_5,\gamma_\mu\}=0$ before trace symmetrization. We employ a version of the ``Bumblebee'' model, where Lorentz symmetry is ostensibly violated, to test these approaches. Using IREG in the physical dimension results in the gap equation consistently evaluating to zero, indicating no Lorentz violation. This is analogous to results from a gauge-invariant sharp cutoff scheme in $4S$, unlike conventional sharp cutoff regularization that depends on symmetric integration and yields different, cutoff-dependent results. However, extending loop momenta and Clifford algebra to quasi-four-dimensional space $Q4S = 4S \oplus X$, while keeping $\gamma_5$ in $4S$, and systematically applying IREG yields finite results comparable to those from the 't Hooft and Veltman (HV) scheme. The $Q4S$ extension offers unique results similar to the BMHV scheme, proving effective when combined with IREG. Despite technically extending beyond the physical dimension, its application remains user-friendly, allowing to perform correctly all integrations in the physical dimension, by appropriately addressing potential symmetry-violating terms through the $X$ space, maintaining otherwise all the established rules of IREG.

A legitimate question emerges as the method delivers consistent results in the sense that ambiguities are absent in the calculations, and yet may break underlying symmetries; this would for instance be the case if used in chiral gauge theories where it is known that the BMHV scheme delivers unique results and yet  may induce spurious anomalies that need be removed by counterterms. With this in mind we have shown in the Appendix C  that the global vector and chiral symmetries can be used as guiding principles (the model is not gauge invariant). We have obtained that promoting the Lagrangian to Q4S the chiral symmetry is violated by an evanescent term, but that this offending term leads to a vanishing contribution for the gap equation. Therefore no counterterms are necessary in this case, as the global chiral symmetry persists in the result of the gap equation in 4S and Q4S.

\vspace{1cm}

\section*{Acknowledgements}

We acknowledge support from Fundação para a Ciência e Tecnologia (FCT) through the projects UIDP/04564/2020\footnote{   https://doi.org/10.54499/UIDB/04564/2020} and UIDB/04564/2020\footnote{https://doi.org/10.54499/UIDP/04564/2020}, and the grant FCT 2020.07172.BD. M. Sampaio acknowledges support from CNPq through grant 302790\slash2020-9. A.C.~ is supported by a postdoctoral fellowship from the Postdoctoral Researcher Program - Resolution GR/Unicamp No. 33/2023.

\appendix

\section{Finite integrals}
\label{finite}

Here we list the finite contributions due to the integrals  $I_2$ and 
$I_2^{\mu\nu}$, eqs.~\ref{eq:I2} and \ref{eq:I2m} respectively. They still have to be integrated over the Feynman parameter $x$, in the final expression \ref{pint}. Defining $A=\int^1_0 dx I_{2fin}$ and $(B+C)^{\mu\nu}=\int^1_0 dx I_{2fin}^{\mu\nu}$, one has
\begin{align}
\label{eq:Is}
    A & =\int^1_0  dx\int_k  \frac{M^4}{(k^2-\mu^2)^2(k^2-M^2-\mu^2)}=\int^1_0  dx M^4 \int^1_0 d\phi \frac{(1-\phi)}{ R(x,\phi)}=\frac{-2 b^2}{9} \left[8 +3 \mbox{ln}\left(\frac{\mu_0}{4}\right)\right]\\ 
      & \nonumber\\
    B^{\mu\nu} &  =\int^1_0  dx \int_k \frac{2 k^\mu k^\nu M^4}{(k^2-\mu^2)^3(k^2-M^2-\mu^2)}\nonumber\\
    &=-\frac{g^{\mu\nu}}{2} \int^1_0  dx M^4   \int^1_0 d\phi \frac{(1-\phi)^2}{ R(x,\phi)}= 
       - g^{\mu\nu} \frac{b^2}{18} \left[19 +6 \mbox{ln}\left(\frac{\mu_0}{4}\right)\right]\\ 
     & \nonumber\\
     C^{\mu\nu} &  =\int^1_0  dx \int_k\frac{k^\mu k^\nu  M^4}{(k^2-\mu^2)^2(k^2-M^2-\mu^2)^2} =\frac{-g^{\mu\nu}}{2}\int^1_0  dx M^4\int^1_0 d\phi \frac{\phi(1-\phi)}{ R(x,\phi)}= \frac{ b^2}{6} g^{\mu\nu}. 
\end{align}

\noindent where $R(x,\phi)=\mu^2+M^2 \phi$, $M^2=4 b^2 x(x-1)$ in terms of the Feynman parameters $x,\phi$. Here $\mu_0=\frac{\mu^2}{b^2}$. In these results the limit $\mu^2\rightarrow 0$ has been taken. Notice the occurrence of infrared divergences in integrals $A$ and $B^{\mu\nu}$, they emerge in the process of separating the $BDI$ from the strictly finite UV contributions in the original integrals $I_2$ and $I_2^{\mu\nu}$, which according to the algorithm \ref{ident} increases the powers of loop momentum in the denominator. As expected these cancel in the final result below \ref{eq:Piap}{\footnote{The parametrization of the infrared divergences adopted is employed successfully in decay and scattering processes in connection with the Kinoshita-Lee-Nauenberg (KLN) theorem in various processes calculated with IREG \cite{gnendiger2017d,pereira2023higgs,torres2021may,rosado2023infrared}}}.  

\noindent Inserting these expressions in eq.~\ref{pint} the final result for the gap equation vanishes in this case
\begin{align}
\label{eq:Piap}
\Pi^\mu=-2(2 b^\mu A -4 b_\nu (B+C)^{\mu\nu})=0
\end{align}


\section{Overview of Implicit Regularization}
\label{sec:ireg}

In section \ref{sec:gamma5} it is explained how the $\gamma_{5}$ can be consistently treated in connection with IREG. The procedure below outlined assumes implicitly that whenever the $\gamma_5$ matrix is present, the operations pertaining to the space $Q4S= 4S \oplus X$ have been performed beforehand.

In this section we present the rules of IREG focusing on one loop order and in the massless limit considered in the present Bumblebee model. A complete $n$-loop set of rules can be found in  \cite{arias2021brief,cherchiglia2021two}.

In IREG, the extraction of the UV divergent content of a Feynman amplitude is done by using algebraic identities at the integrand level. This is done in alignment with Bogoliubov's recursion formula \cite{Bogoliubov:1957gp,Hepp:1966eg,Zimmermann:1969jj}, implying that the way the method defines an UV convergent integral respects locality, Lorentz invariance and unitarity \cite{cherchiglia2011systematic}. IREG has been shown to respect abelian gauge invariance to $n$-loop order \cite{vieira:2015fra, ferreira:2011cv}, as well as non-abelian and SUSY symmetries in specific examples up to two-loop order \cite{batista:2018zxf,cherchiglia2021two, cherchiglia:2015vaa, Fargnoli:2010mf, Carneiro:2003id}. This is achieved in a constrained version of the method, in which surface terms (ST's), which are related to momentum routing of loops in Feynman diagrams, are set to zero. In the realm of applications, processes such as  $h\rightarrow \gamma\gamma$ \cite{Cherchiglia:2012zp}, $e^{-}e^{+}\rightarrow \gamma^{*} \rightarrow q\bar{q}(g)$ \cite{gnendiger2017d}, and $H\rightarrow gg (g)$ \cite{pereira2023higgs} were studied at NLO.

In a nutshell, the rules of IREG are summarized as follows: consider a general $1$-loop Feynman amplitude where we denote by $k$ the internal (loop) momenta, and $p_i$ the external momenta. To this amplitude, we apply the set of rules:

\begin{enumerate}
\item Perform Dirac algebra; 

\item In order to respect numerator/denominator consistency,
as described in the reference \cite{Bruque:2018bmy}, it is necessary to eliminate terms involving internal momenta squared in the numerator by dividing them out from the denominator. For instance,
\begin{equation}
\int_{k}  \dfrac{k^{2}}{k^{2}(k-p)^{2}}\bigg|_{\text{IREG}} \neq g^{\alpha\beta}\int_{k}  \dfrac{k_{\alpha}k_{\beta}}{k^{2}(k-p)^{2}}\bigg|_{\text{IREG}} \quad \mbox{where} \quad \int_k \equiv \int d^4k/(2 \pi)^4.
\end{equation}

\item Include a fictitious mass $\mu^{2}$ in all propagators, where the limit $\mu\rightarrow0$ must be taken at the end of the calculation. In the presence of IR divergences, a logarithm with $\mu^2$ will remain. Assuming that we have an implicit regulator, we apply the following identity in all propagators dependent on the external momenta $p_{i}$ 
	\begin{align}
	\frac{1}{(k-p_{i})^2-\mu^2}=\sum_{j=0}^{n-1}\frac{(-1)^{j}(p_{i}^2-2p_{i} \cdot k)^{j}}{(k^2-\mu^2)^{j+1}}
	+\frac{(-1)^{n}(p_{i}^2-2p_{i} \cdot k)^{n}}{(k^2-\mu^2)^{n}
		\left[(k-p_{i})^2-\mu^2\right]}.
	\label{ident}
	\end{align}
Here $n$ is chosen such that the UV divergent part only has propagators of the form $(k^{2}-\mu^{2})^{-j}$.
\item Express  UV divergencies in terms of Basic Divergent Integrals (BDI's) of the form
\bea
	I_{log}(\mu^2)&\equiv& \int_{k} \frac{1}{(k^2-\mu^2)^{2}},\quad \quad
    I_{log}^{\nu_{1} \cdots \nu_{2r}}(\mu^2)\equiv \int_k \frac{k^{\nu_1}\cdots
		k^{\nu_{2r}}}{(k^2-\mu^2)^{r+2}}.
\eea

\item Surface terms (weighted differences of loop integrals with the same degree of divergence) should be set to zero on the grounds of momentum routing invariance in the loop of Feynman diagrams. This constrained version automatically preserves gauge invariance. For instance,
\bea
\int_k\frac{\partial}{\partial k_{\mu}}\frac{k^{\nu}}{(k^{2}-\mu^{2})^{2}}&=&4\Bigg[\frac{g_{\mu\nu}}{4}I_{log}(\mu^2)-I_{log}^{\mu\nu}(\mu^2)\Bigg]=0.\label{ST1L}
\eea
Similar identities follow for BDI's with a larger number of free Lorentz indexes, as well as for quadratic divergent integrals.

\item A renormalization group scale can be introduced by disentangling the UV/IR behavior of BDI's under the limit $\mu\rightarrow0$. This is achieved by employing the identity
\beq
I_{log}(\mu^2) = I_{log}(\lambda^2) + \frac{i}{(4 \pi)^2} \ln \frac{\lambda^2}{\mu^2},
\label{SR1}
\eeq
It is possible to absorb the BDI's in the renormalisation constants (without explicit evaluation) \cite{Brito:2008zn}, and renormalisation functions can be readily computed using
\beq
\lambda^2\frac{\partial I_{log}(\lambda^2)}{\partial \lambda^2}= -\frac{i}{(4 \pi)^2}.
\eeq

\end{enumerate}

\section{Chiral symmetry}
\label{sec:symmetry}

One of the underlying reasons to propose different regularization methods is to simplify the algebraic calculation (compared to dimensional regularization) at the same time that the symmetries (mostly gauge invariance) of the theory at hand are respected. In this context, the breaking of the symmetry by a regularization method is rather an undesirable aspect than a fundamental reason to dismiss it, since it is always possible (although unpractical) to introduce finite counterterms to enforce the symmetry {\it a posteriori}. The main issue with such approach is when genuine anomalies appear, in which case it is mandatory to separate it from a spurious (regularization induced) anomaly. The identification between spurious and genuine anomalies can be performed by employing the quantum action principle (QAP). Therefore, it is highly desirable that a particular regularization method complies with the QAP. It was proven in~\cite{stockinger2005regularization} that dimensional reduction (from which the HV-regularization is a particular case) does indeed comply with QAP. For IReg, it was shown in~\cite{Bruque:2018bmy} that the necessary conditions to comply with the QAP are fulfilled. However, it is necessary to define the method into Q4S, rather than 4S. Therefore, by employing Q4S-IReg, together with the QAP, one can consistently disentangle a spurious from a genuine anomaly.

For the Bumblebee model considered in this work, gauge symmetry is not present. However, the Lagrangian defined in 4S remains invariant under a global chiral symmetry. As discussed, for instance in \cite{Cherchiglia:2021uce}, it is necessary to define the Lagrangian in Q4S, not 4S, and one has to check if the symmetry is still respected. If this is indeed the case, spurious anomalies will not arrive and no symmetry-restoring counterterm is required. This reasoning is behind the application of the QAP in the BHMV, with the aim to find such countertems, see for instance~\cite{Stockinger:2023ndm}. In the following, we will show that, although the global chiral symmetry is not preserved in Q4S in general, for the gap equation the offending term is not present.

We begin defining the symmetry
\begin{equation}
    \psi\rightarrow e^{i(\alpha+\zeta\gamma^5)}\psi
\end{equation}
where for completeness we consider the combined vector and chiral symmetries (parametrized through $\alpha$ and $\zeta$ respectively).
We recall that $\gamma^{5}$ is defined in 4S, while the Lagrangian given by eq.~\ref{fourferm} is promoted to Q4S, starting from the underlying four fermion interaction in Q4S; it is implicit that the unit matrix multiplying $\alpha$ is defined in Q4S. Using the properties presented in sec.~\ref{sec:reg}, one obtains for an infinitesimal transformation
\begin{equation}
\label{bpsi}
\bar{\psi}\rightarrow \bar{\psi}[(1-i(\alpha-\zeta\gamma_5) -2 i\zeta \gamma_0 \hat{\gamma}_0 \gamma^5]    
\end{equation}
and it is useful to express $\gamma_0 \hat{\gamma}_0$ as the difference of the two unit matrices in Q4S and 4S spaces (thus the last contribution in \ref{bpsi} vanishes). Then
it is possible to show that the Lagrangian is modified as below
\begin{equation}
    \mathcal{L}\rightarrow \mathcal{L} - 2\zeta\bar{\psi}(\gamma^{5}
    \partial_{\mu} + i e B_{\mu}) \hat{\gamma}^\mu  \psi.
\end{equation}

Notice the absence of an $\alpha$- dependent term, thus the vector symmetry is exactly fulfilled in Q4S.

The next step is to derive the effective potential as well as the gap equation. Repeating the reasoning presented in sec.\ref{sec:model}, one obtains
\begin{eqnarray}
\label{gapn}
\frac{dV_{\text{eff}}}{dB_\mu} \Bigg|_{e B_\mu=b_\mu} = &-&\frac{g^2}{e} b^\mu - i e \left[\Pi^\mu + 2\zeta \text{Tr} \int \frac{d^4k}{(2\pi)^4} \frac{1}{(\slashed{k} - \slashed{b}\gamma^5)^{2}} \gamma^\mu \gamma^5(-i\hat{\slashed{k}}\gamma^{5}+ i \slashed{\hat{b}})\right] \nonumber \\
&-&  i e \left[  2\zeta \text{Tr} \int \frac{d^4k}{(2\pi)^4} \frac{1}{(\slashed{k} - \slashed{b}\gamma^5)} i\hat{\gamma}^\mu \right]= 0. \
\end{eqnarray}  
considering terms up to first order in $\zeta$ . One may set the term  proportional to $\hat{b}=b-\bar{b}$ to zero as the vector $b$ is external and only exists in 4S. The other terms containing  $\hat{k}$ and $\hat{\gamma}^\mu$ are due to the breaking of the chiral symmetry in the Lagrangian. If they prove to be non-null, the regularization in Q4S is indeed generating spurious finite terms in the course of the calculation of the gap equation. Such terms should be removed by appropriate symmetry-restoring countertems. In the present case, however, we proceed to show that the offending terms are null.

The rationalized expression for the propagator is given in eq. \ref{prop} and the one for the square of the propagator is 
\begin{equation}
(\slashed{k} - \slashed{b}\gamma^5)^{-2}=\frac{k^2-\bar{b}^2-[\slashed{k},\slashed{\bar{b}}]\gamma^5}{(k-\bar{b})^2(k+\bar{b})^2-4 \hat{k}^2\bar{b}^2}.  
\end{equation}
The term proportional to $\zeta$ in eq.\ref{gapn} is then obtained by analyzing the integrals
\begin{equation}
I^{(1)}_\zeta= \mbox{Tr} \int \frac{d^4k}{(2\pi)^4}\frac{1}{(k-\bar{b})^2(k+\bar{b})^2-4 \hat{k}^2\bar{b}^2}   [(k^2-\bar{b}^2-[\slashed{k},\slashed{\bar{b}}]\gamma^5)\gamma^\mu \gamma^5(-2i\hat{\slashed{k}}\gamma^{5})]   
\end{equation}
corresponding to the integral containing $\hat{k}$
and 
\begin{equation}
I^{(2)}_\zeta= \mbox{Tr} \int \frac{d^4k}{(2\pi)^4}\frac{(k^2+\bar{b}^2+2(k\cdot\bar{b}+[\slashed{k},\slashed{\bar{b}}])\gamma^5)}{(k-\bar{b})^2(k+\bar{b})^2-4 \hat{k}^2\bar{b}^2}[\slashed{k}+\slashed{\bar{b}} \gamma^5](2 i\hat{\gamma}^\mu)  \end{equation}
related to the integral with $\hat{\gamma}^\mu$.  In 
the notation of sec. 4.2 and after taking the trace one obtains
\begin{equation}
I^{(1)}_\zeta= 4\int \frac{d^4k}{(2\pi)^4} \int^1_0 dx \left[\frac{(k^2-2k\cdot\bar{b} (1-2x)+\Delta) \hat{k}^\mu}{(k^2-\Delta)^2} -2\hat{k}_\rho\frac{(k_\delta-\bar{b}_\delta(1-2x)) \bar{b}_\sigma)\epsilon^{\mu\rho\sigma\delta}}{(k^2-\Delta)^2} \right]  
\end{equation} 

\noindent

We recall that $\Delta=4b^2x(x-1)+\mu^2$. It is easy to see that the second integral is identically zero, since it only depends on the vector $b$. Thus, after integration it will be proportional to $\epsilon^{\mu\rho\sigma\delta}b_{\rho}b_{\sigma}b_{\delta} = 0 $, due to the antisymmetric property of $\epsilon^{\mu\rho\sigma\delta}$. For the first integral, we first notice that the terms proportional to $k^{2}$ and $\Delta$ are odd in $k$. Thus, we obtain a null result under integration in the loop momentum. The term proportional to $k\cdot\bar{b}$ can be expressed as  
\begin{align}
I^{(1)}_\zeta&= -8 \int^1_0 dx (1-2x)\bar{b}_{\alpha} \left[\int \frac{d^4k}{(2\pi)^4}\frac{k^{\alpha}k^{\mu}}{(k^2-\Delta)^2}-\bar{g}_{\beta\mu}\int \frac{d^4k}{(2\pi)^4}\frac{k^{\alpha}k^{\beta}}{(k^2-\Delta)^2}\right]\\
&\sim \bar{b}_{\alpha}\left[g^{\alpha\mu}A+b^{\alpha}b^{\mu}B-\bar{g}_{\beta\mu}\left(g^{\alpha\beta}A+b^{\alpha}b^{\beta}B\right)\right]\\
&\sim \left[\bar{b}^{\mu}A+\bar{b}^{2}b^{\mu}B-\bar{b}^{\mu}A-\bar{b}^{2}\bar{b}^{\mu}B\right]
\end{align}
where $A$, $B$ are scalar functions. Since $b$ is an external vector, that in the end of the calculation we return to the 4S space, we obtain that $I^{(1)}_\zeta$ is null. 

As for the integral $I^{(2)}_\zeta$ it reduces to a term identical to the one containing the antisymmetric tensor in  $I^{(1)}_\zeta$ which was proven to be zero and an odd contribution in $k$ which also vanishes upon integration. 

Therefore, no counterterms are necessary, the gap equation respecting the global chiral symmetry in 4S as well as in Q4S.

\addcontentsline{toc}{section}{bibliography}
\bibliographystyle{unsrt}
\bibliography{bibliography}

\end{document}